\documentclass[showpacs,aps,pre]{revtex4}

\usepackage{amssymb}
\usepackage{amsmath}
\usepackage{color}

\makeatletter

\def\boldit#1{\mbox{\boldmath$#1$}}
\def\eqalign#1{\null\vcenter{\def\\{\cr}\openup\jot\m@th
 \ialign{\strut$\displaystyle{##}$\hfil&$\displaystyle{{}##}$\hfil
     \crcr#1\crcr}}\,}

\makeatother

\begin{document}

\def\n{\nabla}

\def\t{ \times }

\def\c{ \cdot }

\def\d{ \text{div} }

\def\rf{ Eq.~(\ref) }

\def\ct{ \cite }

\def\p{ \partial }

\def\pt{ \partial_{t} }

\def\pa{ \partial_{a} }

\def\cu{ \text{curl}}

\def\di{ \text{div}}

\title{Canonical description of ideal magnetohydrodynamic flows and integrals of motion.}

\author{A.~V.~KATS \\
Usikov Institute for Radiophysics and Electronics National Academy of Sciences \\ of Ukraine,
  61085, 12 Ak. Proskury St., Kharkiv, Ukraine \\ Fax: +380 (572) 44-11-05 \;
e-mail:  avkats@akfirst.kharkiv.com; avkats@ire.kharkov.ua}

\date{9.06.03}




\begin{abstract}

In the framework of the variational principle  the canonical
variables describing magnetohydrodynamic (MHD) flows of
general type (i.e., with spatially varying entropy and
nonzero values of all topological invariants) are
introduced. It is shown that the velocity representation of
the Clebsch type following from the variational principle
with constraints is equivalent to that resulting from the
generalization of the Weber transformation performed in the
paper for the case of arbitrary MHD flows. Using such
complete velocity representation enables us not only to
describe the general type flows in terms of single--valued
functions, but also to solve the intriguing problem of the
``missing'' MHD integrals of motion. The  set of hitherto
known MHD local invariants and integrals of motion appears
to be incomplete: for the vanishing magnetic field it does
not reduce to the set of the conventional hydrodynamic
invariants. And if the analogs of the vorticity and helicity
were discussed earlier for the particular cases, the analog
of Ertel invariant has been so far unknown. It is shown that
all ``missing'' invariants are expressed in terms of the
decomposition of the velocity representation into the
``hydrodynamic'' and ``magnetic'' parts. In spite of the
nonunique character of such representation it is shown that
there exists a natural restriction of the gauge
transformations set allowing one to make the invariants
gauge independent. It is found that on the basis of the new
invariants introduced a wide set of high--order invariants
can be constructed. The new invariants are relevant both for
the deeper insight into the problem of the topological
structure of the MHD flows as a whole and for the
examination of the stability problems. The additional
advantage of the proposed approach is that it enables one to
deal with discontinuous flows, including all types of
possible breaks.
\end{abstract}

 \pacs{ 04.20Fy, 47.10.+g, 47.65}

\maketitle


\section{Introduction.}

 \hskip\parindent

It is well--known that description of the solid media flows
in terms of the canonical (Hamiltonian) variables is very
useful and effective, see for instance \cite{Zh_Kuz_97,
GP93}. On the basis of the Hamiltonian variables it is
possible to  deal with  all nonlinear processes in unified
terms which are independent of the specific problem related
to the media under investigation. For instance, all variants
of the perturbation theory are expressed in terms of
different order nonlinear vertices, which along with the
linear dispersion relation contain the specific information
relating to the concrete system, cf. Refs.~\cite{ZLF_92,
KUZ_01}. In studying nonlinear stability  problems the
conventional Hamiltonian approach based upon the
corresponding variational principle allows one to use the
Hamiltonian along with other integrals of motion (momentum,
number of quasi-particles, topological invariants) in order
to construct the relevant Lyapunov functional, cf. Refs.
\cite{Arn65, Abarb_83, Lewis_86, Vl95, Vl96}. Therefore, it
is important to address the problem of introducing the
canonical variables and corresponding variational principle
for the general type MHD flows (i.~e., rotational,
non--barotropic and including all types of breaks that are
possible for MHD) and obtaining the complete set of the
local invariants, see definition and discussions in original
papers \cite{MCTYan_82, STY_90, TYan_93, VTY_95} and in the
recent review \cite{Zh_Kuz_97}. As for the first item, the
example of the variational principle describing all possible
breaks is given in the recent work \cite{KATS_02}.

Here in the framework of some modification of the
variational principle of the latter work we examine the
problem of the MHD invariants. Note  that the set of
invariants for MHD discussed in the literature has been so
far incomplete. The statement becomes apparent when it is
considered that for the vanishing magnetic field this set
has to be reduced to that of the conventional hydrodynamic
invariants. But this limit transition does not reproduce
Ertel, vorticity and helicity invariants existing for the
hydrodynamic flows.

Despite the fact that for the dissipation-free MHD flows there
exist additional topological invariants, namely, magnetic helicity
and cross--helicity, introduced in the papers,
\cite{Woltjer58,SteenbeckKrause66,Mof_69}, the analogs of the
vorticity and helicity invariants have not been discussed with
necessary completeness thus far, cf., for instance, the recent
review \cite{Zh_Kuz_97}. The related quantities were mentioned for
the specific cases of symmetric flows in the works
\cite{Hameiri98, AlmaguerHameiriHerreraHolm88,
IlgisonisPastukhov96}, the vorticity and helicity invariants for
the incompressible flows have been obtained recently in
Refs.~\cite{Vl95, Vl96}. But analog of the Ertel invariant have
not been presented so far (see the short communication in
\cite{KATS_02A}). The problem of obtaining the analogs of the
hydrodynamic invariants consists in the non-potential character of
the Lorentz force. Therefore, the vorticity and helicity of the
total velocity field $\mathbf{v}$ are not conserved along with the
Ertel invariant construction, $\rho^{-1} \boldit{\omega} \c \n s$.
Nevertheless, corresponding generalizations have to exist, which
becomes evident from the simple consideration. Namely, let us
consider the well known set of invariants for dissipation--free
MHD flows (energy, magnetic and cross helicity). Setting the
magnetic field zero we arrive at zero values of the magnetic and
cross helicity invariants, but do not get Ertel invariant (and
hydrodynamic vorticity and helicity for the barotropic flows).
This fact indicates incompleteness of the MHD invariants set.
Evidently, there have to exist MHD analog of the Ertel invariant
passing on to the hydrodynamic Ertel invariant for the vanishing
magnetic field. Below we derive the MHD generalization for the
Ertel invariant and show that the generalized vorticity and
helicity invariants also exist for the compressible barotropic MHD
flows. The possibility of obtaining these invariants is based upon
the velocity decomposition in the two parts, ``hydrodynamic'' and
``magnetic''. The latter vanishes with the magnetic field
vanishing and can be presented in the form of the vector product
of the magnetic field and the canonically conjugate momentum and
was first introduced in the paper \cite{ZAK_KUZ_70}. In spite of
the artificial character of the velocity field decomposition at
first sight, we show that the decomposition naturally follows both
from the least action principle in the canonical variables and
from the partial integration of the Euler equations of motion
(generalized Weber transformation, cf.
Ref.~\cite{Weber1868,lamb}). For the incompressible flows  the
latter was presented in the papers \cite{Vl95,Vl96}. Note that the
``hydrodynamic'' part of the velocity is of the Clebsch type but
involves vector potentials instead of the scalar ones, see
discussion in Refs.~\cite{Zh_Kuz_97,KATS_02,KK_97,KATS_01}. The
use of the vector Clebsch potentials allows one to deal with the
flows possessing nontrivial topology, contrary to the restriction
to the scalar potentials. If the latter are single-valued then the
helicity vanishes identically.

The outline of the paper is as follows. In  section \ref{VAR_PR}
we briefly discuss the appropriate variational principle,
introducing the Clebsch type velocity representation by means of
constraints and defining the canonical variables. The velocity
representation thus obtained  yields the necessary decomposition.
In  section \ref{WEB_TR} we develop generalization of the Weber
transformation and show that it leads to the velocity
representation, which is equivalent to that following from the
variational principle under discussion. In section
\ref{INT_MOTION} we examine the MHD integrals of motion,
introducing ``missing'' MHD invariants, and discuss their
transformation properties relating to the gauge change. We show
that there exist natural gauges under which the additional basic
invariants become unambiguous, specifically that with a vanishing
initial value of the magnetic part of the velocity representation.
In section \ref{Conlusions} we make some conclusions and formulate
problems to be solved later.

\section{Variational principle and canonical variables.}\label{VAR_PR}

 \hskip\parindent

Let us briefly describe the variational principle and subsidiary variables describing
dissipation-free MHD. Starting with the standard Lagrangian density
\begin{equation}\label{Lagr_1A_Er}
  {\mathcal L} = \rho\frac{v^2}{2} -
\rho \varepsilon(\rho, s) + \frac{\mathbf{H}^{2}}{8\pi} \, ,
\end{equation}
where $\rho$, $s$ and $\varepsilon(\rho, s)$ are the fluid density, entropy and internal
energy, respectively, $\mathbf{H}$ denotes the magnetic field, we have to include  the
constraint terms in the action $\mathcal{A}$. Then the action can be presented as
\begin{equation}\label{Action_vol_Er}
\mathcal{A} =  \int d t L' \, , \qquad L' = \int d \mathbf{r} \mathcal{L}'  \, , \qquad
\mathcal{L}' = \mathcal{L} + \mathcal{L}_{c} \, ,
\end{equation}
where $\mathcal{L}_{c}$ is the part of the Lagrangian density respective for the
constraints,
\begin{equation}\label{Constr_8_Er}
 \mathcal{L}_{c} = \rho D\varphi + \boldit{\lambda} D\boldit{\mu} + \sigma D s  -  {\mathbf{M}} \cdot
  \left(
\frac{\partial \mathbf{A}}{\partial t} -  {\mathbf{v}} \t \cu{\mathbf{A}} + \nabla \Lambda  \right) -
\frac{\mathbf{H} \cdot \cu\mathbf{A}}{4 \pi} \, .
\end{equation}
Here $D = \partial_{t} + \mathbf{v} \cdot \nabla$ is the substantial (material) derivative and $\mathbf{A}$ is
the vector potential.\footnote{This form of the action slightly differs from that proposed in Ref.
\cite{KATS_02}. The main difference consists in introducing the vector potential for the magnetic field.
Therefore, here the canonical pair is $\mathbf{A}, -\mathbf{M}$ instead of $\mathbf{H}, \mathbf{S}$, where
$\mathbf{S} = \cu\mathbf{M}$. We do not consider the discontinuous flows and thus we omit the surface term in
the action. But adding corresponding surface term we can easily take the breaks into account.} Including the
terms with $\Lambda$ and $\mathbf{H} = \cu\mathbf{A}$ into $\mathcal{L}_{c}$ allows us to obtain the dynamic
equation for the vector potential in the gauge invariant form (see Eq.~(\ref{Magn_Er}) below)  and to introduce
relation $\mathbf{H} = \cu\mathbf{A}$ strictly into the variational
principle. 

Supposing first that all variables introduced (including velocity) are independent, we
obtain the following set of variational equations
\begin{equation}\label{mass_Er}
\delta \varphi \Longrightarrow \quad \partial_{t} \rho + \di(\rho{\mathbf{v}}) = 0  ,
\end{equation}
\begin{equation}\label{VOL1B_Er}
\delta \rho  \Longrightarrow \quad  D\varphi =  w  - v^2/2  \, ,
    \end{equation}
\begin{equation}\label{mu_1_Er}
\delta  \boldit{\lambda} \Longrightarrow \quad   D \boldit{\mu} = 0 \, ,
\end{equation}
  \begin{equation}\label{lambda_Er}
\delta  \mu_{m}   \Longrightarrow \quad       \partial_{t} \lambda_{m} + \di(\lambda_{m} \mathbf{v}) =0  ,
   \end{equation}
\begin{equation}\label{s_eq_Er}
\delta  \sigma \Longrightarrow \quad  D s = 0 ,
\end{equation}
\begin{equation}\label{entr_Er}
\delta  s \Longrightarrow \quad     \partial_{t} \sigma + \di(\sigma\mathbf{v}) =   - \rho T ,
   \end{equation}
\begin{equation}\label{Magn_Er}
\delta  \mathbf{M}  \Longrightarrow \quad   \partial_{t}  \mathbf{A} =   \mathbf{v} \t \cu\mathbf{A} - \nabla
\Lambda  ,
\end{equation}
  \begin{equation}\label{VOL1A_Er}
\delta   \mathbf{A}  \Longrightarrow \quad   \partial_{t}  \mathbf{M}  = \frac{\cu\mathbf{H}}{4\pi} +
\cu[\mathbf{v} \t \mathbf{M}]  .
      \end{equation}
\begin{equation}\label{DIV_Er}
\delta   \mathbf{H} \Longrightarrow \quad   \mathbf{H} = \cu\mathbf{A} ,
\end{equation}
\begin{equation}\label{Er_1}
\delta  \Lambda \Longrightarrow \quad \di\mathbf{M} = 0 ,
\end{equation}
where $w$ and $T$ are the enthalpy density and
temperature.\footnote{Note that substituting $\mathbf{M} = \cu
\mathbf{S}$ into Eq. (11) one can integrate it and arrive at the
dynamic equation for $\mathbf{S}$, $\pt \mathbf{S} = (4\pi)^{-1}
\mathbf{H} + \mathbf{v} \t \cu \mathbf{S} + \n \Psi$, where $\Psi$
represents a scalar field respectful for the $\mathbf{S}$ gauge.
This relation only by the  $\mathbf{S}$ sign differs from equation
(10.9) of the reference \cite{Zh_Kuz_97} (or Eq.~(7) in the
original paper \cite{ZAK_KUZ_70}). }

Note that in this section we assume the velocity field to be independent of other
variables. Therefore, variation with respect to $\mathbf{v}$ results in the velocity
representation
\begin{equation}\label{velocity_Er}
\delta   \mathbf{v}   \Longrightarrow \quad  \rho {\bf v} = - \rho \nabla \varphi - {\lambda}_{m} \nabla
{\mu}_{m} - \sigma \nabla s -  {\bf H} \t \mathbf{M}  .
\end{equation}
It is convenient to rewrite it in a shortened form that emphasizes its structure. Bearing in mind that the
velocity potential $\varphi$, the vector Lagrange markers $\boldit{\mu}$, entropy $s$  and the vector potential
$\mathbf{A}$ can be treated as generalized coordinates, one can see that $\rho$, $\boldit{\lambda}$, $\sigma$
and subsidiary field $-\mathbf{M}$ are conjugated momenta, respectively. Let\footnote{We do not include
$\Lambda$ into the set of canonical variables dealing with the extended Hamiltonian description, cf.
Refs.~\cite{Dirac64, Gitman86}. Otherwise, we can include $\Lambda$ into the set of generalized coordinates.
Denoting corresponding conjugate momentum $\pi_{\Lambda}$ and adding to the Hamiltonian density,
Eq.~(\ref{30_06_02D}), the term $\pi_{\Lambda}\nu$ results in
    $$
\partial_{t} \pi_{\Lambda} = - \delta \mathcal{H}/\delta \Lambda = - \di\mathbf{M} ,
\quad \partial_{t} \Lambda = \delta \mathcal{H}/\delta \pi_{\Lambda} = \nu .
    $$
Variation of the action with respect to the additional variable $\nu$ results now in the restriction
$\pi_{\Lambda} = 0$. This restriction is consistent with the set of canonical equations. Namely,
Eq.~(\ref{VOL1A_Er}) (or, equivalently, the canonical equation $\partial_{t}  \mathbf{M} =  \delta H/\delta
\mathbf{A}$) leads to $\partial_{t} \di\mathbf{M} = 0$.  The momentum $\pi_{\Lambda}$ is the linear function of
$t$, $\pi_{\Lambda} = \pi_{\Lambda}(t_{0}) - (t - t_{0}) \di\mathbf{M}$. The condition $\pi_{\Lambda} = 0$
follows for the specific choice $\pi_{\Lambda}(t_{0}) = 0$ and $\di\mathbf{M}(t_{0}) = 0$. Moreover, as it
becomes clear below, the initial condition $\mathbf{M}(t_{0}) = 0$ (resulting in $\di\mathbf{M}(t_{0}) =
\di\mathbf{M}(t) = 0$) leads to essential simplifications. On the other hand, conditions $\pi_{\Lambda} = 0$,
$\mathbf{M}(t_{0}) = 0$ do not lead to any restrictions on the physical variables and the MHD flow as well. The
subsidiary functions $\Lambda$ and $\nu$ can be expressed in terms of other variables as $\Lambda = \Delta^{-1}
(\d [\mathbf{v} \t \mathbf{H}] - \d \dot{\mathbf{A}}) + \Lambda'$, $\nu = \partial_{t}\Lambda$, where $\Delta$
denotes the Laplace operator and $\Lambda'$ is arbitrary solution of the Laplace equation.}\label{footn_2}

\begin{equation}\label{5_09_02}
\mathcal{ Q} = (Q, \mathbf{A}) , \quad  Q = (\varphi, \boldit{\mu}, s) , \quad \mathcal{ P} = \delta
\mathcal{A}/\delta (\partial_{t}\mathcal{ Q}) , \quad  P = (\rho, \boldit{\lambda}, \sigma) , \quad \mathcal{ P}
= (P , -\mathbf{M}) .
\end{equation}

Then the velocity representation takes the transparent form
\begin{equation}\label{5_09_02_1}
\mathbf{v} = \mathbf{v}_{0}(\mathcal{P}, \nabla \mathcal{Q}) , \quad \mathbf{v}_{0} = \mathbf{v}_{h} +
\mathbf{v}_{M} \, , \quad  \mathbf{v}_{h} = - \frac{P}{\rho} \nabla Q , \quad \mathbf{v}_{M} = - \frac{{\bf H}
\t \mathbf{M}}{\rho}  .
\end{equation}
Here subindexes $h$ and $M$ correspond to the ``hydrodynamic'' and
``magnetic'' parts of the velocity field. The suffix zero on
$\mathbf{v}$ underlines the fact that $\mathbf{v}_{0}$ is supposed
to be the dependent variable, it is expressed in terms of the
canonical variables by means of the representation found. The
hydrodynamic part, $\mathbf{v}_{h}$, corresponds to the
generalized Clebsch representation, cf. \cite{KATS_02, KK_97,
KATS_01}, and the magnetic part, $\mathbf{v}_{M}$, coincides with
the traditional term if we replace the divergence-free field
$\mathbf{M}$ by $\cu\mathbf{S}$. This term was first introduced by
Zakharov and Kuznetsov, cf. Ref.~\cite{ZAK_KUZ_70}. But they used
the incomplete form for the hydrodynamic part of the velocity,
restricting it to the scalar Clebsch variables. This reduced
form\footnote{Note that the considerations based upon the Pfaff's
theorem also result in the reduced velocity representation. But
this theorem in our case claims only {\underline{the local
equivalence}} between the three-dimensional vector field and the
standard form $\phi + \lambda \n \mu$ with the appropriate scalars
$\phi$, $\lambda$, $\mu$. This point is often
ignored.}\label{footn_3} evidently corresponds to the flows with
zero--valued generalized helicity (or, for $\mathbf{H} \rightarrow
0$, it results in the hydrodynamic helicity vanishing) if  the
scalar Clebsch variables are single--valued. Besides, the above
velocity representation involves the entropy term, $-\sigma \n
S/\rho$. The latter is essential for the discontinuous flows with
any types of the dynamically allowable breaks, including shocks,
cf. Refs.~\cite{KK_97, KATS_01, KATS_02, KATS_02A}. Note that
Lagrange markers $\boldit{\mu}$ are continuous crossing the break
surface, contrary to the entropy. Therefore, the entropy term can
be omitted for the continuous flows when the entropy can be
considered as a continuous function depending on the Lagrange
markers.

From the velocity representation Eq.~(\ref{5_09_02_1}) and the equations of motion
(\ref{mass_Er}) -- (\ref{VOL1A_Er}) it strictly follows that the velocity field
$\mathbf{v} = \mathbf{v}_{0}$ satisfies Euler equation with the magnetic force taken into
account. Namely, providing differentiation we have
\begin{equation}\label{Euler_Er}
  \rho D \mathbf{v}_{0} = - \nabla p + \frac{\cu \mathbf{H} \t \mathbf{H}}{4\pi}  ,
\end{equation}
where $p$ is the fluid pressure.

\subsection{Canonical variables.}\label{CAN_VAR}

 \hskip\parindent

The variational principle can be easily reformulated in the Hamiltonian form. Excluding
the magnetic and velocity fields by means of Eqs.~(\ref{DIV_Er}), (\ref{5_09_02_1}) we
arrive at the following Hamiltonian density
\begin{equation}\label{30_06_02D}
 \mathcal{H} = \mathcal{H}(\mathcal{P}, \nabla \mathcal{Q}) = \mathcal{P} \partial_{t}  \mathcal{Q} -  \mathcal{L}' =  \rho\frac{v_{0}^2}{2} +
\rho \varepsilon(\rho, s) + \frac{(\cu\mathbf{A})^{2}}{8\pi} - \mathbf{M} \cdot \nabla \Lambda  .
\end{equation}
Equations of motion (\ref{mass_Er}) -- (\ref{VOL1A_Er}) can now be expressed in the
canonical form
\begin{equation}\label{30_06_07}
\partial_{t} {\cal Q} = {\delta {\cal H}}/{\delta { \cal P}} \, , \qquad
\partial_{t} {\cal P} = - {\delta {\cal H}}/ {\delta { \cal Q}}  \, ,
\qquad  {\cal Q} = (\varphi, \boldit{\mu}, s; \mathbf{A} ) \, , \quad  {\cal P} = (\rho, \boldit{\lambda},
\sigma ; - \mathbf{M} ) \, ;
\end{equation}
Eq.~(\ref{DIV_Er}) serves as a definition of the magnetic field, and the divergence--free
condition for the subsidiary field $\mathbf{M}$, Eq.~(\ref{Er_1}), follows from the
variation of the action
\begin{equation}\label{Action_vol_Er_1}
\mathcal{A} =  \int d t  \int d \mathbf{r} \left( \mathcal{P}\partial_{t} \mathcal{Q} -
\mathcal{H}\right)
\end{equation}
with respect to $\Lambda$. Note that it is possible to put $\Lambda = 0$. Under this assumption the
divergence--free condition for the field $\mathbf{M}$ vanishes. But from Eq.~(\ref{VOL1A_Er}) it follows that
$\di\mathbf{M}$ is a conserved quantity, $
\partial_{t} \di\mathbf{M} = 0$. Therefore, supposing that $\di\mathbf{M} = 0$ holds
for some initial moment we arrive at the conclusion that this is valid for the arbitrary
moment. Nevertheless, it proves convenient to deal with $\Lambda \ne 0$ that makes it
possible to use different gauge conditions for the vector potential.

The above variational principle results in the set of dynamic equations. From the latter follow the conventional
MHD equations, (\ref{mass_Er}), (\ref{entr_Er}), (\ref{Euler_Er}) and the equation for the magnetic field, which
follows from Eq.~(\ref{Magn_Er}) after taking curl,
\begin{equation}\label{5_09_02_2}
\partial_{t}  \mathbf{H} =   \cu[\mathbf{v} \t \mathbf{H}] .
\end{equation}

On the contrary, if at some initial moment, $t = \bar{t}$, we have the conventional MHD fields $\bar{\rho}$,
$\bar{s}$, $\bar{\mathbf{v}}$ and  $\overline{\mathbf{H}}$, then we can find the initial subsidiary fields
$\bar{\varphi}$, $\bar{\boldit{\mu}}$, $\bar{\boldit{\lambda}}$, $\bar{\sigma}$, $\bar{\mathbf{A}}$,
$\overline{\mathbf{M}}$ and $\bar{\Lambda}$, satisfying Eqs.~(\ref{DIV_Er}) -- (\ref{velocity_Er}). This can be
done to within the gauge transformations (the latter do not change both the velocity and the magnetic field) due
to the fact that the subsidiary fields play a role of generalized potentials. Then, if the uniqueness conditions
are satisfied both for the conventional MHD equations and for the set of variational equations, we are led to
conclude that corresponding solutions coincide for all moments. In this sense we can state that these sets of
equations are equivalent, cf. Ref.~\cite{KUZ_01}.

The complete representation of the velocity field in the form of the generalized Clebsch representation
(\ref{5_09_02_1}) allows, first, to deal with the MHD flows of general type, including all types of breaks, cf.
Ref.~\cite{KATS_02}; second, for the zero magnetic field it results in the correct limit transition to the
conventional hydrodynamics, cf. Refs.~\cite{KATS_01,KK_97}; third, it allows obtaining the integrals and
invariants of motion for the MHD flows additional to the known ones: for instance, the generalized Ertel
invariant, generalized vorticity and generalized helicity, see below. The two last integrals were deduced for
the particular case of incompressible flows in the papers \cite{Vl95, Vl96}, cf. also papers \cite{Hameiri98,
AlmaguerHameiriHerreraHolm88, IlgisonisPastukhov96} where the vorticity and helicity analogs were obtained for
the MHD flows with the specific spatial symmetry.

Moreover, it is possible to show that representation (\ref{5_09_02_1}) is equivalent to that following from the
Weber transformation, cf. Refs.~\cite{Weber1868,lamb} and the recent review \cite{Zh_Kuz_97}.

\section{Generalized Weber transformation.}\label{WEB_TR}

\hskip\parindent

Suppose that the fluid particles are labelled by Lagrange markers $\mathbf{a} = (a_{1},
a_{2},a_{3})$. The label of the particle passing through point $\mathbf{r} = (x_{1},
x_{2},x_{3})$ at time $t$ is then
\begin{equation}\label{W_1}
\mathbf{a} = \mathbf{a}(\mathbf{r}, t)  , \quad  D \mathbf{a} = \frac{\partial \mathbf{a}}{\partial t} +
(\mathbf{v} \cdot \nabla )\mathbf{a} = 0 .
\end{equation}
The particle paths and velocities are given by the inverse function
\begin{equation}\label{W_3}
\mathbf{r} = \mathbf{r}(\mathbf{a} , t)   , \quad \mathbf{v} = D \mathbf{r} (\mathbf{a} ,
  t) =  \left. \partial \mathbf{r}/\partial t \right|_{\mathbf{a} = const}  .
\end{equation}
Let the initial position of the particle labelled $\mathbf{a}$ is $\mathbf{X}$, i.e.,
\begin{equation}\label{W_4}
 \mathbf{r}(\mathbf{a}, 0) = \mathbf{X}(\mathbf{a})   .
\end{equation}
A natural choice of the labels would be $\mathbf{X}(\mathbf{a}) = \mathbf{a}$; however it
is convenient to retain the extra freedom represented by the ``rearrangement function''
$\mathbf{X}(\mathbf{a})$.

We seek to transform the equation of motion (\ref{Euler_Er}) to an integrable form, by generalizing the argument
of Weber \cite{Weber1868}  (see, for example, Refs.~\cite{Serrin59}, \cite{Zh_Kuz_97}, and \cite{Vl95}. It is
convenient to represent Eq.~(\ref{Euler_Er}) as
\begin{equation}\label{Euler_W}
  D \mathbf{v} = - \nabla w + T \nabla s +  \mathbf{j} \t \mathbf{h}   ,
\end{equation}
where  $\mathbf{h} = \mathbf{H} /\rho$ and the vector $\mathbf{j}$ is defined according to
\begin{equation}\label{Current}
  \mathbf{j} = \frac{\cu\mathbf{H}}{4\pi}  ,
\end{equation}
being proportional to the current density. Multiplying Eq.~(\ref{Euler_W}) by $\partial
x_{k}/ \partial a_{i}$ we have
\begin{equation}\label{W_8}
(D v_{k}) \frac{\partial x_{k}}{\partial a_{i}} = - \frac{\partial w}{\partial x_{k}} \frac{\partial
x_{k}}{\partial a_{i}} + T \frac{\partial s}{\partial x_{k}} \frac{\partial x_{k}}{\partial a_{i}} +  [
{\mathbf{j}} \t {\mathbf{h}} ]_{k} \frac{\partial x_{k}}{\partial a_{i}} \, .
\end{equation}
The l.h.s. can be represented as
\begin{equation}\label{W_9}
(D v_{k}) \frac{\partial x_{k}}{\partial a_{i}} = D \left(v_{k} \frac{\partial
x_{k}}{\partial a_{i}}\right) - \frac{\partial }{\partial a_{i}} (v^{2}/2) ,
\end{equation}
where  we have taken into account that  operator $D \equiv
\partial /\partial t |_{{\mathbf{a}}= const}$ and therefore $D x_{k} =
v_{k}$ and $D$ commutes with derivative $\partial / \partial a_{i}$. Eq.~(\ref{W_8}) now
takes  the form
\begin{equation}\label{31_07_02_8}
D \left(v_{k} \frac{\partial x_{k}}{\partial a_{i}}\right)  = \frac{\partial }{\partial a_{i}} (v^{2}/2 - w) + T
\frac{\partial s}{\partial a_{i}} + [ {\mathbf{j}} \t {\mathbf{h}} ]_{k} \frac{\partial x_{k}}{\partial a_{i}}
\, .
\end{equation}
It is convenient to transform the last term  by means of the dynamic equation for the
subsidiary field $\mathbf{m} = \mathbf{M}/\rho$ (compare Eq.~(\ref{VOL1A_Er}))
\begin{equation}\label{31_07_02_9}
D \mathbf{m}   =  (\mathbf{m} \cdot  \nabla ) \mathbf{v} + \mathbf{j}/ \rho  .
\end{equation}
Then we can transform the last term in the r.h.s. of Eq.~(\ref{31_07_02_8}) to the form
of the substantial derivative, see Appendix\label{???},
\begin{equation}\label{31_07_02_11}
[ {\mathbf{j}} \t {\mathbf{h}} ]_{k} \frac{\partial x_{k}}{\partial a_{i}} =  D \left( [ {\mathbf{m}} \t
{\mathbf{H}} ]_{k}  \frac{\partial x_{k}}{\partial a_{i}} \right) .
\end{equation}

Analogously, the first two terms in the r.h.s. of Eq.~(\ref{31_07_02_8}) can be presented
as substantial derivatives by means of introducing subsidiary functions $\varphi$ and
$\sigma$, which satisfy equations (compare Eqs.~(\ref{entr_Er}), (\ref{VOL1B_Er}))
\begin{equation}\label{31_07_02_13}
D \left( \frac{\sigma}{\rho} \right)  = - T ,
\end{equation}
\begin{equation}\label{31_07_02_15}
D \varphi =  w - v^{2}/2 .
\end{equation}
Then
\begin{equation}\label{31_07_02_14}
T \frac{\partial s}{\partial a_{i}} = - \frac{\partial s}{\partial a_{i}} D \left(
\frac{\sigma}{\rho} \right) = - D \left( \frac{\partial s}{\partial
a_{i}}\frac{\sigma}{\rho} \right) , \quad \frac{\partial }{\partial a_{i}} (v^{2}/2 - w)
= - D \left( \frac{\partial  \varphi }{\partial a_{i}} \right) ,
\end{equation}
where we have taken into account that $D s = 0$ along with $D (\partial s / \partial
a_{i}) = 0$. Therefore, we can present the Euler equation (\ref{31_07_02_8}) in the
integrable form
\begin{equation}\label{31_07_02_17}
D \left(v_{k} \frac{\partial x_{k}}{\partial a_{i}}\right)  = - D \left( \frac{\partial \varphi }{\partial
a_{i}} \right) - D \left( \frac{\partial s}{\partial a_{i}}\frac{\sigma}{\rho} \right) + D \left( [ {\mathbf{m}}
\t {\mathbf{H}} ]_{k} \frac{\partial x_{k}}{\partial a_{i}} \right) .
\end{equation}
Integration leads to the relation
\begin{equation}\label{31_07_02_18}
v_{k} \frac{\partial x_{k}}{\partial a_{i}}  = -  \frac{\partial  \varphi }{\partial a_{i}}  - \frac{\partial
s}{\partial a_{i}}\frac{\sigma}{\rho}  - [\mathbf{H} \t \mathbf{m}]_{k} \frac{\partial x_{k}}{\partial a_{i}} +
b_{i} ,
\end{equation}
Here $\mathbf{b} = \mathbf{b}(\mathbf{a})$ does not depend on time explicitly, $D
\mathbf{b} = 0$, presenting the vector constant of integration. Multiplying this relation
by $\partial a_{i} /\partial x_{j}$ allows  reverting from Lagrangian $(\mathbf{a}, t)$,
to the Eulerian,  $(\mathbf{r}, t)$, variables,
\begin{equation}\label{31_07_02_20}
\mathbf{v} = - \nabla \varphi + b_{k} \nabla  a_{k} -  \frac{\sigma}{\rho} \nabla s - \mathbf{h} \t \mathbf{M} .
\end{equation}

This representation obviously coincides with the above discussed Clebsch representation
if one identifies $\mathbf{b}$ with $-\boldit{\lambda}/\rho$ and $\mathbf{a}$ with
$\boldit{\mu}$. Moreover, this proves equivalence of description of the general--type
magnetohydrodynamic flows  in terms of canonical variables introduced and the
conventional description in Lagrange or Euler variables. The equations of motion for the
generalized coordinates and momenta follow now from definitions of the subsidiary
variables $\mathbf{a}$, $\mathbf{m} = \mathbf{M}/\rho$, $\sigma$, $\varphi$ and
$\mathbf{b}$.

Emphasize  that the vector field $\mathbf{M} = \rho \mathbf{m}$ introduced by
Eq.~(\ref{31_07_02_9}) satisfies the integral  relation
\begin{equation}\label{12_08_02_D_Er}
\partial_{t}  \int_{\Sigma} \mathbf{M} \cdot   d \boldit{\Sigma} = \int_{\Sigma} \mathbf{j} \cdot   d
\boldit{\Sigma} ,
\end{equation}
where $\boldit{\Sigma}$ is some oriented area moving with the
fluid. This fact was first indicated  in Ref.~\cite{Vl95} for the
incompressible flows. Now we see that it holds true for the
general case. The proof of this statement is given in Appendix.
Expressing $\mathbf{M} = \cu\mathbf{S}$ and making use of the
Stokes theorem we conclude that time derivative of the vector
$\mathbf{S}$ circulation over the closed frozen--in contour
$\partial \Sigma$ is proportional to the current (recall,
$\mathbf{j} = (4\pi)^{-1} \cu\mathbf{H}$ and differs from the
current density by the constant multiplier) intersecting the
surface defined by this contour,
\begin{equation}\label{12_08_02_D5_Er}
\partial_{t}  \int_{\partial \Sigma} \mathbf{S} \cdot   d \mathbf{l} = \int_{\Sigma} \mathbf{j} \cdot   d
\boldit{\Sigma} = (4\pi)^{-1} \int_{\partial \Sigma} \mathbf{H} \cdot   d \mathbf{l}
\end{equation}
that highlights the physical meaning of the subsidiary field
$\mathbf{S}$ usually introduced for the canonical description of
MHD flows. Underline that this identity strictly follows from the
dynamic equation for the subsidiary field $\mathbf{S}$ and is
insensitive to the compressibility.

The vector constant of integration, $\mathbf{b}$,  may be expressed in terms of the
initial conditions,
\begin{equation}\label{31_07_02_21}
\begin{split}
b_{i} = \overline{V}_{k} (\mathbf{a}) \frac{\partial X_{k}}{\partial a_{i}}  +
\frac{\partial \varphi_{0} }{\partial a_{i}}
+ c_{0} \frac{\partial s}{\partial a_{i}} \, , \quad \quad \quad \quad \quad  \quad \quad  \quad \quad \\
\varphi_{0} = \varphi(\mathbf{a}, 0) , \quad c_{0} = \left( \frac{\sigma}{\rho} \right)
\biggl|_{t = 0}  ,  \quad \overline{V}_{k} (\mathbf{a}) = V_{k}
(\mathbf{a}) + [\mathbf{h}_{0} \t \mathbf{M}_{0}]_{k} \, , \quad {V}_{k} (\mathbf{a}) = v_{k} (\mathbf{a}, 0) \, , \quad\quad \\
\mathbf{h}_{0} \equiv \mathbf{h}_{0}(\mathbf{a}) = \mathbf{h} (\mathbf{x}(\mathbf{a}, 0),
0) = \mathbf{h} (\mathbf{X}(\mathbf{a}) , 0) ,  \quad \mathbf{M}_{0} \equiv
\mathbf{M}_{0}(\mathbf{a}) = \mathbf{M} (\mathbf{x}(\mathbf{a}, 0), 0)  = \mathbf{M}
(\mathbf{X}(\mathbf{a}), 0) .
\end{split}
\end{equation}

Under special conditions, namely, for
\begin{equation}\label{1_08_02_1}
\mathbf{X}(\mathbf{a}) =  \mathbf{a} , \quad \mathbf{r}(\mathbf{a}, 0) = \mathbf{a}  ,
\quad \mathbf{a}(\mathbf{r}, 0) = \mathbf{r} ,
\end{equation}
from Eq.~(\ref{31_07_02_21}) it follows
\begin{equation}\label{1_08_02_2}
b_{i} = \overline{V}_{i} (\mathbf{a})  + \frac{\partial  \varphi_{0} }{\partial a_{i}} +
c_{0} \frac{\partial s}{\partial a_{i}} \, .
\end{equation}
Adopting zero initial conditions,
\begin{equation}\label{1_08_02_3}
\mathbf{M}_{0}= 0 , \quad \varphi_{0} = 0  , \quad \sigma_{0} = 0  ,
\end{equation}
we obtain
\begin{equation}\label{2_08_02}
\mathbf{b} = \overline{\mathbf{V}}(\mathbf{a}) =  \widetilde{\mathbf{v}}(\mathbf{a}, 0)
\equiv  \widetilde{\mathbf{v}}_{0}(\mathbf{a}) = \mathbf{v}(\mathbf{a}, 0) \, ,
\end{equation}
where symbol $\,{\widetilde{}}\,$   indicates that we are dealing with the velocity field
in the Lagrange description, i.e., $\widetilde{\mathbf{v}}(\mathbf{a}, t)$ denotes the
velocity of the fluid particle with label $\mathbf{a}$ at time $t$. Evidently,
$\widetilde{\mathbf{v}}(\mathbf{a}, t) = \mathbf{v}(\mathbf{r}, t)$, where $\mathbf{a}$
and $\mathbf{r}$ are linked by relations (\ref{W_1}) and (\ref{W_3})  for the specific
choice given by Eqs.~(\ref{1_08_02_1}), (\ref{1_08_02_3}). Then the velocity
representation takes the particular form
\begin{equation}\label{2_08_02_1}
\mathbf{v} = \mathbf{v}_{h} -  [\mathbf{h} \t \mathbf{M}] , \quad \mathbf{v}_{h} \equiv - \nabla \varphi +
\widetilde{v}_{0k} \nabla  a_{k} -  \frac{\sigma}{\rho} \nabla s , \quad \widetilde{\mathbf{v}}_{0}(\mathbf{r})
= \mathbf{v}(\mathbf{r}, 0) , \quad \mathbf{a}(\mathbf{r} , 0) = \mathbf{r} .
\end{equation}
It differs from that presented in Ref.~\cite{KUZ_01} by involving the entropy term. Note that existence of this
term allows one to describe the general--type MHD flows (and hydrodynamic flows under condition $\mathbf{H} =
0$) with arbitrary possible discontinuities, including shocks, slides and rotational breaks, cf.
Ref.~\cite{KK_97, KATS_01, KATS_02}. One can omit this term for continuous barotropic and isentropic flows.

\section{Integrals of motion.}\label{INT_MOTION}

\hskip\parindent

The conservation laws, as is well-known, follow from the specific symmetries of the action. Existence of the
relabelling transformations group (first discussed by Salmon in Ref. \cite{Salmon82}) of the Lagrange markers,
$\boldit{\mu}$, leads to the integrals of motion that are additional to the energy, the fluid momentum and mass
conservation. These additional integrals are expressed in terms of the Lagrange description of the motion, i.e.,
in terms of the Lagrange markers, etc. Therefore, as a rule, they are gauge--dependent. The frozen--in character
of the magnetic field results in the specific topological integrals of motion, namely, magnetic helicity and
cross--helicity, first discussed  in Refs.~\cite{Woltjer58,SteenbeckKrause66,Mof_69},  see also review
\cite{Zh_Kuz_97}. Corresponding densities are respectively
\begin{equation}\label{8_09_02}
h_{M} = \mathbf{A \cdot   H} ,
\end{equation}
and
\begin{equation}\label{8_09_02_1}
h_{C} = \mathbf{v \cdot   H} .
\end{equation}

To clarify the following discussion relating to the additional
local invariants and integrals of motion, let us briefly recall
the known ones.  As it strictly follows from the dynamic
equations, the local conservation law for the magnetic helicity
holds true for general type MHD flows
\begin{equation}\label{8_09_02_2}
\partial_{t} h_{M} + \di \mathbf{q}_{M} = 0  , \quad \mathbf{q}_{M} =
\mathbf{v} h_{M} - \mathbf{H} \cdot \left( \mathbf{A}  \cdot   \mathbf{v}  - \Lambda \right) .
\end{equation}
On the contrary, in the general case the cross-helicity is governed by equation
    $$
\partial_{t} h_{C} /  \partial t = - \di\left[\mathbf{v}  h_{C} +  ( w - v^{2}/2 ) \mathbf{H}
\right] + T \di(s \mathbf{H})
    $$
and is not conserved. But for barotropic and isentropic flows the
pressure $p = p(\rho)$ and $h_{C}$ is conserved,
\begin{equation}\label{8_09_02_3}
\partial_{t} h_{C} + \di\mathbf{q}_{C} = 0 ,  \quad \mathbf{q}_{C} =
\mathbf{v}  h_{C} +  ( \chi - v^{2}/2 ) \mathbf{H} ,
\end{equation}
where $\chi = \int d p/\rho$.

For the general case  one more conserved quantity first discovered by Gordin and
Petviashvili, cf. Ref.~\cite{Gordin87},  is known. Corresponding density is
\begin{equation}\label{9_09_02}
h_{P} = \mathbf{H} \cdot   \nabla s ,
\end{equation}
and
\begin{equation}\label{9_09_02_1}
\partial_{t} h_{P} + \di\mathbf{q}_{P} = 0 ,  \quad \mathbf{q}_{P} = \mathbf{v} h_{P} \, .
\end{equation}

The integral conservation laws are related to the local conserved quantities. For
instance, integrating $h_{P}$ over arbitrary substantial volume $\widetilde{V}$ we obtain
conserved quantity $\mathcal{I}_{P}$,
\begin{equation}\label{9_09_02_2}
\mathcal{I}_{P} = \int_{\widetilde{V}} d \mathbf{r} h_{P} \, , \quad \partial_{t}
\mathcal{I}_{P} = 0 .
\end{equation}

Note that $h_{P}/\rho$ gives us an example of the so--called local
Lagrange invariants (in other words, Casimirs), cf.
Refs.~\cite{MCTYan_82, STY_90, TYan_93, VTY_95} and \cite{GP93,
Zh_Kuz_97}. By definition they obey the following equations
\begin{equation}\label{8_07_02_Er}
\partial_{t} \alpha + \mathbf{v} \cdot   \nabla  \alpha = 0 \, , \quad \partial_{t} {\mathbf{I}} + (\mathbf{v} \cdot
\nabla ) \mathbf{I} = 0 \, ,
\end{equation}
\begin{equation}\label{8_07_02_2_Er}
\partial_{t} {\mathbf{J}} + (\mathbf{v} \cdot   \nabla ) \mathbf{J} - (\mathbf{J} \cdot   \nabla ) \mathbf{v}  =
0 \, ,
\end{equation}
\begin{equation}\label{8_07_02_1_Er}
\partial_{t} {\mathbf{L}} + (\mathbf{v} \cdot   \nabla ) \mathbf{L} + (\mathbf{L} \cdot   \nabla ) \mathbf{v} +
\mathbf{L} \t \cu\mathbf{v} = 0  , \; {\mbox{or, equivalently,}}  \; \;
\partial_{t} {\mathbf{L}} + \nabla (\mathbf{v} \cdot    \mathbf{L}) - \mathbf{v} \t
\cu\mathbf{L} = 0  .
\end{equation}
Here $\alpha$ and $\mathbf{I}$ denote the scalar and vector
Lagrange invariants, $\mathbf{J}$ is the frozen--in field, and
$\mathbf{L}$ denotes $S$-type invariant by terminology of
Ref.~\cite{TYan_93}, related to a frozen--in surface. To these
invariants it is necessary to add the density $\rho$.\footnote{In
terms of the differential forms $\alpha$ and $\mathbf{I}$ are
scalar and vector $0$--forms; $\mathbf{L}$, $\mathbf{J}$  and
$\rho$ are $1$--, $2$-- and $3$--forms, respectively.
}\label{foot_d} Evidently, the quantity $h_{P}/\rho$ is the
$\alpha$-type invariant. The Lagrange markers $\boldit{\mu}$ and
quantities $\boldit{\lambda}/\rho$ supply examples of the vector
Lagrange invariants, the magnetic field $\mathbf{H}$ divided by
$\rho$, $\mathbf{h} = \mathbf{H}/\rho$, is invariant of the
$\mathbf{J}$- type, gradient of any scalar Lagrange invariant is
the $S$-type invariant,
\begin{equation}\label{9_09_02_5}
\mathbf{L}' = \nabla \alpha .
\end{equation}

There also exist other relations between different type invariants, see Refs.~\cite{GP93,
Zh_Kuz_97}, allowing one to generate new invariants. For instance, the scalar product of
the $\mathbf{J}$ and $\mathbf{L}$ invariants results in some scalar Lagrange invariant,
symbolically
\begin{equation}\label{9_09_02_6}
\alpha' = (\mathbf{J} \cdot   \mathbf{L}) .
\end{equation}
The above mentioned invariant $h_{P}/\rho$ can be obtained by means of this relation if
we put $\mathbf{J} = \mathbf{h}$ and $\mathbf{L} = \nabla s$. Other examples are
represented by relations generating $\mathbf{J}$- ($\mathbf{L}$)- type invariants by
means of two $\mathbf{L}$- ($\mathbf{J}$-) type invariants,
\begin{equation}\label{9_09_02_6A}
\mathbf{J}' = [\mathbf{L} \t \mathbf{L}']/\rho ,
\end{equation}
\begin{equation}\label{28_09_02_1}
\mathbf{L}' = \rho [\mathbf{J} \t \mathbf{J}'] .
\end{equation}

Note that integrating of the density $h_{M}$ over an arbitrary substantial volume does
not lead to the conserved integral. It is easy to check up that
\begin{equation}\label{9_09_02_3}
\mathcal{I}_{M} = \int_{\widetilde{V}} d \mathbf{r} h_{M}
\end{equation}
satisfies
\begin{equation}\label{9_09_02_4}
\partial_{t} \mathcal{I}_{M} = \int_{\partial \widetilde{V}} d \Sigma
\left( \mathbf{A}  \cdot   \mathbf{v}  - \Lambda \right) H_{n}    \, ,  \quad H_{n} = \mathbf{H}  \cdot
\mathbf{n} \, ,
\end{equation}
where integration in the r.h.s. is performed over  the boundary $\partial \widetilde{V}$
of the volume $\widetilde{V}$, $\mathbf{n}$ is the outward normal and $d \Sigma$ denotes
an infinitesimal area of the surface $\partial \widetilde{V}$. It is obvious that
$\mathcal{I}_{M}$ will be an integral of motion if $H_{n}$ equals zero. This fact allows
us to conclude that $\mathcal{I}_{M}$ becomes an integral of motion if we choose the
substantial volume in such a way that on the boundary of the initial volume,
$\widetilde{V}|_{t = t_{0}}$, holds equality $H_{n}|_{t = t_{0}} =0$. The latter
condition is invariant of the motion: if equality $H_{n} = 0$ is fulfilled for the
initial moment, then it holds true in the future.

Another way to make $\mathcal{I}_{M}$ invariant consists in fixing the gauge of the vector potential
$\mathbf{A}$ so that $\mathbf{A}  \cdot \mathbf{v}  = \Lambda$. Then the dynamic equation for $\mathbf{A}$,
(\ref{Magn_Er}), takes  the form
    $$
\partial_{t} \mathbf{A}
+ \nabla (\mathbf{v} \cdot   \mathbf{A}) - \mathbf{v} \t \cu \mathbf{A} = 0 ,
    $$
i.e., $\mathbf{A}$ becomes an invariant of the $\mathbf{L}$-- type. Under this gauge
condition the quantity $h_{M}/\rho$ presents the scalar Lagrange invariant, $D
(h_{M}/\rho) = 0$.

As for the local conservation law for the cross-helicity, Eq.~(\ref{8_09_02_3}), it
obviously leads to the integral conserved quantity $\mathcal{I}_{C}$ for the barotropic
flows but with the following restriction: integration has to be performed over the
specific substantial volume, namely such that condition $H_{n}|_{\partial \widetilde{V}}
= 0$ (this condition is invariant of the motion) holds,
    $$
\partial_{t} \mathcal{I}_{C} = 0 \,  , \quad \mathcal{I}_{C} \equiv  \int_{\widetilde{V}} d \mathbf{r}
h_{C} \, , \quad H_{n}|_{\partial \widetilde{V}} = 0 .
    $$

Existence of the recursive procedure allowing one to construct new invariants on the
basis of the starting set of invariants, see Refs.~\cite{GP93, Zh_Kuz_97}, underlines the
role of the local invariants among other conserved quantities. Although in terms of the
Lagrangian variables (such as the markers $\boldit{\mu}$) there exists a wide set of
invariants, see, for instance, Ref.~\cite{Zh_Kuz_97}, the most interesting invariants are
such that can be expressed in Eulerian (physical) variables and thus are
gauge--invariant.

Emphasize that in the conventional hydrodynamics there exists Ertel invariant
$\alpha_{E}$,
\begin{equation}\label{11_09_02}
 \alpha_{E} = h_{E}/\rho , \quad h_{E} = \boldit{\omega} \cdot  \nabla s ,
\end{equation}
where $\boldit{\omega} = \cu\mathbf{v}$ is vorticity,
\begin{equation}\label{11_09_02_1}
\partial_{t} h_{E} + \di\mathbf{q}_{E} = 0 ,  \quad \mathbf{q}_{E} =  h_{E}\mathbf{v} , \quad D \alpha_{E} = 0
.
\end{equation}
The corresponding integral of motion reads
\begin{equation}\label{12_09_02}
\partial_{t} \mathcal{I}_{E} =  0 \,  ,
\quad \mathcal{I}_{E} \equiv  \int_{\widetilde{V}} d \mathbf{r} h_{E} \, .
\end{equation}
Note that $D \mathcal{I}_{E} = 0$ holds true for an arbitrary substantial volume $
\widetilde{V}$.

The Ertel invariant density has the structure of the Eq.~(\ref{9_09_02_6}) with
$\mathbf{L} = \nabla s$, $\mathbf{J} = \boldit{\omega}/\rho$ (recall that
$\boldit{\omega}$ is a frozen-in field for the barotropic hydrodynamic flows). In the
hydrodynamic case there also exists the helicity invariant
\begin{equation}\label{12_09_02_1}
h_{H} = \boldit{\omega} \cdot  \mathbf{v} ,
\end{equation}
which has a topological meaning, defining  knottedness of the flow. It satisfies equation
\begin{equation}\label{12_09_02_2}
\partial_{t} h_{H} + \di\mathbf{q}_{H} = 0 , \quad \mathbf{q}_{H} = h_{H} \mathbf{v} + (\chi -
v^{2}/2) \boldit{\omega}  ,
\end{equation}
and evidently results in  the corresponding integral conservation law
\begin{equation}\label{12_09_02_3}
\partial_{t} \mathcal{I}_{H} =  0 \,  ,
\quad  {\mbox{for}} \quad  \omega_{n}|_{\partial \widetilde{V}} = 0 \, , \quad
\mathcal{I}_{H} \equiv  \int_{\widetilde{V}} d \mathbf{r} h_{H} \,  .
\end{equation}

For the MHD case the vector $\boldit{\omega}/\rho$ is not the frozen--in field due to the
fact that magnetic force is non--potential. It seems evident that for the MHD case there
have to exist the integrals of motion generalizing the conventional helicity and Ertel
invariant along with vorticity integral. These invariants are to pass into the
conventional ones for the vanishing  magnetic field. The generalization for the vorticity
and helicity invariants was obtained in the paper \cite{Vl95} for the particular case of
the incompressible flows. In the following section it is shown that there exists MHD
generalization for the Ertel invariant, and results of the paper \cite{Vl95} relating to
the vorticity and helicity can be extended to incompressible barotropic MHD flows.

\subsection{Generalized vorticity.}\label{Gen_vorticity}

\hskip\parindent

Let us prove that the quantity $\boldit{\omega}_{h}/\rho$, where
\begin{equation}\label{12_09_02_4}
\boldit{\omega}_{h} \equiv \cu\mathbf{v}_{h}  = - \left[\nabla \left(\frac{P}{\rho}\right) \t \nabla Q\right] =
- \left[\nabla \left(\frac{\lambda_{m}}{\rho}\right) \t \nabla \mu_{m}\right] - \left[\nabla
\left(\frac{\sigma}{\rho}\right) \t \nabla s\right] ,
\end{equation}
is the frozen--in field  (``hydrodynamic'' part of the vorticity) for the barotropic MHD flows. It would be a
trivial consequence of the fact that $[\mathbf{L} \t \mathbf{L}']/\rho$, where $\mathbf{L}$, $\mathbf{L}'$ are
Lamb type invariants, is the local invariant of the frozen--in type if all quantities $Q$ and $P/\rho$ satisfy
homogeneous transport equations being $\alpha$- or $\mathbf{I}$- type invariants (remember that $\nabla \alpha$
and $\nabla I_{m}$ are $\mathbf{L}$--type invariants). But $\varphi$ and $\sigma/\rho$ satisfy the inhomogeneous
equations of motion. Therefore, let us start with equation of motion for the ``hydrodynamic'' part of the
velocity. Differentiating (\ref{5_09_02_1}) and making use of relations
    $$
D (\nabla X) = \nabla (D  X) - (\nabla v_{m}) \cdot \partial_{m} X
    $$
we have
    $$
D \mathbf{v}_{h} = - D \left(\frac{P}{\rho}\right) \cdot \nabla Q - \frac{P}{\rho} \cdot \nabla (D Q) +
\frac{P}{\rho} (\nabla v_{m}) \cdot \partial_{m} Q = T \nabla s - \nabla (w - v^{2}/2) - v_{hm}\nabla v_{m} \, ,
    $$
or, after simple rearrangements,
\begin{equation}\label{12_09_02_5}
D \mathbf{v}_{h} = - \nabla p /\rho + (v_{m} - v_{hm}) \cdot \nabla v_{m}  \, .
\end{equation}
Taking the curl of this equation results in
    $$
\partial_{t} \boldit{\omega}_{h} = - \cu(v_{m} \partial_{m}  \mathbf{v}_{h}) + [\nabla \rho \t \nabla
p]/\rho^{2} - \cu(v_{hm} \nabla v_{m}) =
    $$
    $$
= [\nabla \rho \t \nabla p]/\rho^{2} + \cu(v_{m} \nabla v_{hm} - v_{hm} \nabla v_{m}) \, .
    $$
The term in the square brackets is equal to $v_{m} \nabla v_{hm} - v_{hm} \nabla v_{m} = \mathbf{v} \t
\boldit{\omega}_{h}$ and we obtain
\begin{equation}\label{12_09_02_6}
\partial_{t} \boldit{\omega}_{h} = [\nabla \rho \t \nabla p]/\rho^{2}  + \cu [\mathbf{v} \t
\boldit{\omega}_{h}] \, .
\end{equation}
For the barotropic flows, $p = p(\rho)$, the first term in the r.h.s. becomes zero and we can see that
$\boldit{\omega}_{h}/\rho$ is the frozen--in field,
\begin{equation}\label{12_09_02_6A}
D \left(\frac{\boldit{\omega}_{h}}{\rho} \right) = \left(\frac{\boldit{\omega}_{h}}{\rho}
 \cdot  \nabla \right)\mathbf{v} \, .
\end{equation}
At $\mathbf{H} = 0$, $\boldit{\omega}_{h}$ corresponds to the conventional hydrodynamic
vorticity.

\subsection{Generalized Kelvin's theorem.} \label{Generalized Kelvin}


The frozen--in character of the generalized vorticity allows obtaining the strict generalization of the Kelvin's
theorem for the barotropic flows. But with some restrictions it is valid also for the non barotropic flows.
Namely, circulation $\Gamma$ of the hydrodynamic part of the velocity over the closed material contour
$\mathcal{C}$ is a constant of motion if the entropy $s$ is constant on this contour,
\begin{equation}\label{29_12_02_1A}
D \Gamma = 0 , \quad \Gamma \equiv \oint_{\mathcal{C}} \mathbf{v}_{h} \cdot d\mathbf{l} \quad \text{for} \quad
s|_{{C}} = const .
\end{equation}
The proof strictly follows from the velocity representation:
\begin{equation}\label{29_12_02_2A}
\Gamma = \oint_{\mathcal{C}} \mathbf{v}_{h} \cdot d\mathbf{l} =
\oint_{\mathcal{C}} \left( d \varphi + \frac{ \lambda_{m}}{\rho} d
\mu_{m} + \frac{\sigma}{\rho} d s\right) = \oint_{\mathcal{C}}
\left(\frac{\lambda_{m}}{\rho} d \mu_{m} + \frac{\sigma}{\rho} d
s\right)  .
\end{equation}
Differentiating $\Gamma $ and taking into account that $D \mu_{m} = D (\lambda_{m}/\rho)
= 0$ we obtain
\begin{equation}\label{29_12_02_3A}
D \Gamma  = \oint_{\mathcal{C}} d s D \left( \frac{\sigma}{\rho} \right)  = - \oint_{\mathcal{C}} T d s = 0
\quad \text{for} \quad   s|_{\mathcal{C}} = const.
\end{equation}

Note that for barotropic flows this result strictly follows from the fact that
$\boldit{\omega}_{h}/\rho$ is the frozen--in field. Namely, for any  $\mathbf{J}$--type
invariant  it can be easily proved that
    $$
D \int_{\Sigma} d \Sigma \, \rho \mathbf{J}  \cdot \mathbf{n}  = 0 ,
    $$
where integration is performed over the substantial surface $\Sigma$. Then for $\rho
\mathbf{J} = \boldit{\omega}_{h}$ after applying the Stokes theorem, we have
   $$
D \int_{\Sigma} d \Sigma \, \boldit{\omega}_{h}  \cdot \mathbf{n} =  D \oint_{\partial \Sigma} \mathbf{v}_{h}
\cdot d\mathbf{l} = 0 .
    $$


\subsection{Generalized helicity.}\label{Gen_helicity}

\hskip\parindent

Now it can be proved that generalized helicity, $h_{H}$, defined in terms of the ``hydrodynamic''  part of the
velocity,
\begin{equation}\label{13_09_02}
h_{H} = \boldit{\omega}_{h}  \cdot  \mathbf{v}_{h} ,
\end{equation}
is the integral of motion for barotropic flows. Differentiating Eq.~(\ref{13_09_02}) and
taking into account Eqs.~(\ref{12_09_02_5}), (\ref{12_09_02_6}) for barotropic flows we
arrive at the local conservation law of the form (rather cumbersome calculations are
given in Appendix):
\begin{equation}\label{13_09_02_1}
\partial_{t} h_{H} + \di\mathbf{q}_{H} = 0 , \quad \mathbf{q}_{H} =
h_{H}\mathbf{v} + (\chi - v^{2}/2)\boldit{\omega}_{h}   \, .
\end{equation}
In analogy with the hydrodynamic case we can conclude that the integral helicity
$\mathcal{I}_{H}$ (defined  by means of Eq.~(\ref{12_09_02_3})) is the integral
invariant, moving together with the fluid if the normal component of the vorticity tends
to zero, $\omega_{hn} = 0$, on the surface of the corresponding substantial volume
$\widetilde{V}$. Note that the condition $\omega_{hn} = 0$ is invariant of the motion
(due to the frozen--in character of $\boldit{\omega}_{h}/\rho$) and therefore it can be
related to the initial surface only.

\subsection{Generalized Ertel invariant.}


Let us show that there exists strict generalization of the Ertel invariant for the MHD
case. For this purpose let us prove that without any restrictions related to the
character of the flow the quantity
\begin{equation}\label{13_09_02_2}
h_{E} = (\boldit{\omega}_{h} \cdot  \nabla s)
\end{equation}
obeys the conservation law of the form
\begin{equation}\label{15_09_02}
\partial_{t} h_{E} + \di\mathbf{q}_{E} = 0 , \quad \mathbf{q}_{E} = h_{E} \mathbf{v} .
\end{equation}
Equivalently, the quantity $\alpha_{E} = h_{E}/\rho$ is transported by the fluid
\begin{equation}\label{15_09_02_1}
D  \alpha_{E} = 0 , \quad \alpha_{E} = h_{E}/\rho ,
\end{equation}
being $\alpha$- type invariant. For the barotropic flows it immediately follows from the
fact that $\boldit{\omega}_{h}/\rho$ is the frozen--in field if the composition rules
given by Eqs.~(\ref{9_09_02_6}) and (\ref{9_09_02_5}) are taken into account. In order to
make the proof for the non barotropic flows more transparent let us consider a more
general situation. Let $\widetilde{\mathbf{J}}$ satisfy equation of motion of the form
\begin{equation}\label{15_09_02_2}
D  \widetilde{\mathbf{J}}  =  (\widetilde{\mathbf{J}} \cdot  \nabla) \mathbf{v} +
\mathbf{Z} ,
\end{equation}
differing from the frozen field equation (\ref{8_07_02_2_Er}) by existence of the term
$\mathbf{Z}$ that violates homogeneity. Then, if $\alpha$ represents any scalar Lagrange
invariant, we have
    $$
D  (\widetilde{\mathbf{J}}  \cdot  \nabla \alpha) =  D \widetilde{\mathbf{J}} \cdot \nabla \alpha  +
\widetilde{\mathbf{J}}  \cdot  D \nabla \alpha
 =     \mathbf{Z}  \cdot  \nabla \alpha
+  \left((\widetilde{\mathbf{J}}  \cdot  \nabla)  \mathbf{v} \right)  \cdot  \nabla \alpha  -
(\widetilde{\mathbf{J}} \cdot  \nabla v_{m}) \cdot \partial_{m} \alpha   .
$$
Here the two last terms cancel and we get
\begin{equation}\label{15_09_02_3}
D  (\widetilde{\mathbf{J}}  \cdot  \nabla \alpha )  =     \mathbf{Z}  \cdot  \nabla \alpha  \quad {\mbox{if}}
\quad  D  \widetilde{\mathbf{J}}  = (\widetilde{\mathbf{J}} \cdot  \nabla) \mathbf{v} + \mathbf{Z} \quad
{\mbox{and}} \quad D \alpha = 0 .
\end{equation}
For $\mathbf{Z} = 0$ these relations prove the generating rule of Eq.~(\ref{9_09_02_6}). But we can see that
$\widetilde{\mathbf{J}}  \cdot  \nabla \alpha $ becomes the local Lagrange invariant under less restrictive
condition $\mathbf{Z}  \cdot  \nabla \alpha = 0$. That is the case for the Ertel invariant: $\mathbf{Z} =
[\nabla \rho \t \nabla p]/\rho^{3}$ is orthogonal to $\nabla s$ due to the fact that the scalar product of any
three thermodynamic quantities is equal to zero (because any thermodynamic variable in the equilibrium state is
a function of two basic variables). This concludes the proof.

The conserved integral quantity associated with $\alpha_{E}$ is
\begin{equation}\label{15_09_02_4}
\mathcal{I}_{E} = \int_{ \widetilde{V}} d \mathbf{r} h_{E} \, ,  \quad
\partial_{t} \mathcal{I}_{E} = 0  .
\end{equation}
Note that by the structure $\mathcal{I}_{E}$ is not gauge--invariant in contrast to the
hydrodynamic case. Let us examine its change under gauge transformation that results in
$\mathbf{v}_{h} \Rightarrow \mathbf{v}'_{h}$, $\mathbf{v}_{M} \Rightarrow
\mathbf{v}'_{M}$ with
    $$
\mathbf{v}'_{h} + \mathbf{v}'_{M} = \mathbf{v}_{h} + \mathbf{v}_{M} \, .
    $$
Then
    $$
\mathcal{I}'_{E} - \mathcal{I}_{E}  = \int_{ \widetilde{V}}  d \mathbf{r} \, \nabla s \cdot
(\boldit{\omega}'_{h} - \boldit{\omega}_{h}) =  \int_{ \widetilde{V}} d \mathbf{r} \, \nabla s \cdot
(\boldit{\omega}_{M} - \boldit{\omega}'_{M}) \, .
    $$
But $\nabla s \cdot  (\boldit{\omega}_{M} - \boldit{\omega}'_{M}) = - \di[\nabla s \t (\mathbf{v}'_{M} -
\mathbf{v}_{M})]$ and, therefore,
    $$
\mathcal{I}'_{E} - \mathcal{I}_{E}  =   - \int_{ \partial \widetilde{V}} d \Sigma \,\mathbf{n} \cdot  [\nabla s
\t (\mathbf{v}'_{M} - \mathbf{v}_{M})]  \, .
    $$
Now we can proceed in the two ways. First, making use of identity $\nabla s \t \mathbf{X} = \cu (s \mathbf{X}) -
s \cdot \cu \mathbf{X}$ we obtain
    $$
\mathcal{I}'_{E} - \mathcal{I}_{E}  =   - \int_{ \partial \widetilde{V}} d \Sigma \, \mathbf{n}\cdot \left(
\cu(s (\mathbf{v}'_{M} - \mathbf{v}_{M})) - s \, \cu (\mathbf{v}'_{M} - \mathbf{v}_{M}) \right) \, .
    $$
Here the integral of the first term vanishes (that is trivial for a closed boundary
$\partial \widetilde{V}$ and assumes the necessary decrease of the integrand for the
infinite volume $ \widetilde{V}$) and we have
\begin{equation}\label{16_12_02}
\mathcal{I}'_{E} - \mathcal{I}_{E}  =    \int_{ \partial \widetilde{V}} d \Sigma \, s \mathbf{n} \cdot
\cu(\mathbf{v}'_{M} - \mathbf{v}_{M})  \, .
\end{equation}
This representation immediately suggests that integral Ertel invariant becomes
gauge--independent for the substantial volume $\widetilde{V}$ chosen in such a way that
its boundary coincides with the entropy--constant surface, $s|_{\partial \widetilde{V}} =
const$.

The second way is as follows. Bearing in mind that $\mathbf{v}'_{M} - \mathbf{v}_{M} = - [\mathbf{h} \t
(\mathbf{M}' - \mathbf{M}) ]$ we obtain
\begin{equation}\label{15_09_02_5}
\mathcal{I}'_{E} - \mathcal{I}_{E}  = \int_{ \partial \widetilde{V}} d \Sigma \, \mathbf{n} \cdot  [\nabla s \t
[\mathbf{h} \t (\mathbf{M}' - \mathbf{M}) ]]  .
\end{equation}
Inasmuch as both $\mathbf{M}'$ and $\mathbf{M}$ satisfy Eq.~(\ref{VOL1A_Er}), their
difference, $\overline{\mathbf{M}} = \mathbf{M}' - \mathbf{M}$,  is governed by the
homogeneous equation
    $$
 \partial_{t}  \overline{\mathbf{M}}  = \cu [\mathbf{v} \t \overline{\mathbf{M}}]  ,
    $$
i.e. $\overline{\mathbf{m}} = \overline{\mathbf{M}}/\rho$ is
frozen--in field. Then we can conclude that the vector $\nabla s
\t [\mathbf{h} \t \overline{\mathbf{m}}] $ entering the integrand
is the frozen--in field, as it follows from recursion relations
(\ref{9_09_02_5}) -- (\ref{9_09_02_6A}). Therefore, if we adopt
relation $\mathbf{n} \cdot  [\nabla s \t [\mathbf{h} \t
\overline{\mathbf{m}}]]|_{\partial \widetilde{V}} = 0$ as the
initial condition, then it holds true for all moments. For
instance, this relation is fulfilled if $H_{n}=0$ and
$\overline{m}_{n} = 0$ at the initial moment. Evidently, these two
conditions cannot be fulfilled for an arbitrary gauge. But we can
restrict ourselves to a such subset of the initial conditions for
the subsidiary field $\mathbf{M}$ that $\mathbf{M}|_{t = t_{0}} =
f \mathbf{H}|_{t = t_{0}}$, where $f$ is an arbitrary function.
(Then $\di\mathbf{M} = \di\mathbf{M}|_{t = t_{0}} = \nabla f \cdot
\mathbf{H}|_{t = t_{0}}$ is time--independent in accordance with
equation (\ref{VOL1A_Er}), and for the particular choice of $f$
such  that $\mathbf{H}|_{t = t_{0}} \cdot \nabla f  =0 $ we have
$\di\mathbf{M} = 0$.) For these initial conditions $\mathbf{M}$
along with $\mathbf{M}'$ are collinear to $\mathbf{H}$ at the
initial moment and therefore the initial value of the scalar
product $\mathbf{n} \cdot [\nabla s \t [\mathbf{h} \
\overline{\mathbf{m}}]]$ is zero. Due to the frozen--in character
of the quantity $\nabla s \t [\mathbf{h} \t
\overline{\mathbf{m}}]$ equation $\mathbf{n} \cdot [\nabla s \t
[\mathbf{h} \t \overline{\mathbf{m}}]] = 0$ holds true for the
arbitrary moment. Thus we can make the conclusion that gauge
dependence of the Ertel's invariant can be partly eliminated by
appropriate choice of the initial conditions or substantial
volumes.

\subsection{Specific gauge.}

Examination of the integrals of motion shows that they are
gauge--dependent. This dependence is attributed to different
decompositions of the velocity field into the ``hydrodynamic'' and
``magnetic'' parts, $\mathbf{v} = \mathbf{v}_{h} +
\mathbf{v}_{M}$. Underline that there exists a wide subgroup of
the gauge transformations that include transformations that change
the generalized potentials $\varphi$, $\boldit{\mu}$,
$\boldit{\lambda}$ and $\sigma$ with no change in $\mathbf{M}$.
The ``hydrodynamic'' part of the velocity representation,
$\mathbf{v}_{h}$, does not evidently vary under the action of
transformations of this, say, ``hydrodynamic'' subgroup. Then the
generalized circulation, integral helicity and Ertel integral are
likewise invariant under these gauge transformations. The simplest
way to restrict the gauge transformations by this subgroup
consists in adopting zero initial conditions for the subsidiary
field $\mathbf{M}$. This choice does not restrict in any way the
character of a flow, in particular, all integrals of motion can
possess nonzero values. The more detailed discussion of the gauge
dependence of the additional integrals considered will be
presented elsewhere.


\section{Conclusions.}\label{Conlusions}


The results obtained can be summarized as follows. First,  the variant of introducing the
canonical description of the MHD flows by means of the variational principle with
constraints is presented. It is shown that in order to describe general--type MHD flows
it is necessary to use in the generalized Clebsch--type representation for the fluid
velocity field the vector Clebsch variables (the Lagrange markers and conjugate momenta)
along with the entropy term (cf. papers \cite{KK_97, KATS_01} describing the hydrodynamic
case) and the conventional magnetic term introduced first in Ref.~\cite{ZAK_KUZ_70}. Such
a complete representation allows one to deal with general--type MHD flows, including all
type of breaks, see Ref. \cite{KATS_02}. Second, it is proved that the generalized Weber
transformation introduced leads to the velocity representation, which is equivalent to
that introduced by means of the variational principle. Third, the  existence of the
generalized Ertel invariant for MHD flows  is proven. Forth, there are generalized the
vorticity and helicity invariants for the compressible barotropic MHD flows (first
discussed for the incompressible case in cf. \cite{Vl95}). Fifth, the relations between
the local and integral invariants are discussed along with the gauge dependence of the
latter.

As a consequence of the completeness of the proposed velocity representation we get the
correct limit transition from the MHD to the conventional hydrodynamic flows. The results
obtained allow one to consider the complicated MHD problems in terms of the Hamiltonian
variables. The use of this approach was demonstrated for the specific case of
incompressible flows in the series of papers \cite{Vl95, Vl96}  devoted to the nonlinear
stability criteria. Emphasize that existence of the additional invariants proved in our
paper is of high importance for the stability problems.

Note that existence of the additional basic invariants of the
motion makes it actual to examine the problem of the complete set
of independent invariants, and, respectively, the complete set of
the corresponding Casimirs,  cf. Ref.~\cite{Zh_Kuz_97}. For
instance, existence of the three independent basic local
invariants for the non barotropic flows ($s$, $\alpha_{E}$ and
$\mathbf{h}$) immediately leads to the two denumerable sets of the
monomial scalar invariants
$$
\alpha_{P}^{(m)} = (\mathbf{h} \cdot \nabla)^{m} s  , \quad \alpha_{E}^{(m)} =
(\mathbf{h} \cdot \nabla)^{m} \alpha_{E} \, , \quad \alpha_{P}^{(1)} = \alpha_{P} \, ,
\quad \alpha_{E}^{(0)} = \alpha_{E} \, , \quad m =  0, 1, 2, \ldots
$$
The first set was discussed in the paper \cite{Zh_Kuz_97}, and the second subset is a new one along with the
``parent'' Ertel invariant $\alpha_{E}$. Evidently,
    $$
\widetilde{\alpha} = f\left( \{ \alpha_{P}^{(m)}\} ,  \{ \alpha_{E}^{(m')} \} \right) ,
    $$
where $f$ is an arbitrary function, is also the scalar Lagrange
invariant. Therefore, we immediately arrive at the following set
of integrals of motion (Casimirs)
\begin{equation}\label{9_09_03}
\mathcal{I} = \int_{\tilde{V}} d \mathbf{r} \rho f\left( \{
\alpha_{P}^{(m)}\} ,  \{ \alpha_{E}^{(m')}\}  \right) ,
\end{equation}
which is much  wider than that discussed in the literature, cf.
Eq.~(10.23) in Ref.~\cite{Zh_Kuz_97}. The additional set of the
scalar monomial Lagrange invariants can be generated by the
magnetic helicity under the specific gauge condition, $\Lambda =
\mathbf{A} \c \mathbf{v}$, $\alpha_{M}^{(n)} = (\mathbf{h}\c
\n)^{n} \alpha_{M}$. This enables evident generalization of the
integrals of motion (\ref{9_09_03}).

One example of the additional $\mathbf{J}$-- invariants reads
    $$
\mathbf{J}' = [\nabla s \t \nabla  \alpha_{E}]/\rho .
    $$
In turn,  one can get new sets of the scalar invariants by
applying operation $( \mathbf{J}' \cdot \nabla) $ to the previous
scalar invariants and so on. Obviously, this also leads to
additional Casimirs to that indicated in (\ref{9_09_03}).

For the barotropic flows the picture is analogous: the basic set
of the scalar Lagrange invariants involves the generalized
helicity, $\alpha_{H} = h_{H}/\rho$, and $\alpha_{M}$ (under the
gauge condition specified above), and   with
$\mathbf{J}$-invariants $\mathbf{h}$ and $\boldit{\omega}_{h}$.
Therefore, we obtain additional scalar invariants
$\alpha_{H}^{(n)} = (\mathbf{h}\c \n)^{n} \alpha_{H}$,
$\alpha_{M}^{(n)}$, $\widetilde{\alpha}_{H}^{(n)} = (\rho^{-1}
\boldit{\omega}_{h}\c \n)^{n} \alpha_{H}$,
$\widetilde{\alpha}_{M}^{(n)} = (\rho^{-1} \boldit{\omega}_{h}\c
\n)^{n} \alpha_{M}$ and  the conserved integrals of the form
\begin{equation}\label{9_09_03_1}
\mathcal{I} = \int_{\tilde{V}} d \mathbf{r} \rho f\left( \{
\alpha_{H}^{(n)}\} ,  \{ \alpha_{M}^{(n')}\} ,
 \{ \widetilde{\alpha}_{H}^{(n'')}\} ,  \{
\widetilde{\alpha}_{M}^{(n''')}\} \right) .
\end{equation}
This set of the Casimirs generalizes that presented in
cf.~Ref.~\cite{Zh_Kuz_97}, the latter follows from
(\ref{9_09_03_1}) if we replace the function $f$ depending on the
four sets of the monomial invariants by a function depending only
on the invariants $\alpha_{M}^{(n)}$.

Note that we can construct the following generations of the local
invariants by means of the recursion relations and obtain Casimirs
of a more sophisticated structure than that presented in
(\ref{9_09_03}), (\ref{9_09_03_1}). The problem of obtaining the
complete set of the local invariants and gauge invariance of the
corresponding integral invariants is rather complicated and is
still under examination. This questions will be discussed in
detail in the forthcoming paper.

\section*{Appendix A}


In order to prove Eq.~(\ref{31_07_02_11}) let us substitute $\mathbf{j}$ from Eq.~(\ref{31_07_02_9}) into
expression $[ {\mathbf{j}} \t {\mathbf{h}} ]_{k} {\partial x_{k}}/{\partial a_{i}}$. Then
\begin{equation}\label{31_07_02_10EA}
\begin{split}
[ {\mathbf{j}} \t {\mathbf{h}} ]_{k} \frac{\partial x_{k}}{\partial a_{i}} = [ D{\mathbf{m}} \t {\mathbf{H}}
]_{k} \frac{\partial x_{k}}{\partial a_{i}} - [ (\mathbf{m} \cdot \nabla ) \mathbf{v} \t {\mathbf{H}} ]_{k}
\frac{\partial
x_{k}}{\partial a_{i}} = \\
= \frac{\partial x_{k}}{\partial a_{i}} D \left( [ {\mathbf{m}} \t {\mathbf{H}} ]_{k} \right) -  \left( [
{\mathbf{m}} \t D ( \rho \mathbf{h}) ]_{k} + [ (\mathbf{m} \cdot \nabla ) \mathbf{v} \t {\mathbf{H}} ]_{k}
\right) \frac{\partial x_{k}}{\partial a_{i}} .
\end{split}
\end{equation}
Proceeding with the terms in the second brackets we obtain
  \begin{equation}\label{28_09_02}
\begin{split}
[ {\mathbf{m}} \t D ( \rho \mathbf{h}) ]_{k} + [ (\mathbf{m} \cdot  \nabla ) \mathbf{v} \t {\mathbf{H}} ]_{k} =
[ {\mathbf{m}} \t \mathbf{h} ]_{k} \cdot D  \rho + [ \rho \mathbf{m} \t D   \mathbf{h} ]_{k}  + [ (\mathbf{m}
\cdot \nabla ) \mathbf{v} \t {\mathbf{H}} ]_{k} =
\\ =  - [ \mathbf{M} \t \mathbf{h} ]_{k} \cdot \di\mathbf{v} + [ \mathbf{M} \t (\mathbf{h}
 \cdot  \nabla) \mathbf{v} ]_{k}  + [ (\mathbf{M} \cdot  \nabla ) \mathbf{v} \t {\mathbf{h}} ]_{k}  =
- [\mathbf{M} \t \mathbf{h}]_{s}
\partial_{k} v_{s} \, ,
\end{split}
\end{equation}
where $\mathbf{M} = \rho \mathbf{m}$ and the dynamic equation $D \mathbf{h} = (\mathbf{h} \cdot  \nabla )
\mathbf{v}$ along with identity
$$
[ \mathbf{M} \t (\mathbf{h}  \cdot  \nabla) \mathbf{v} ]_{k} + [ (\mathbf{M} \cdot  \nabla ) \mathbf{v} \t
{\mathbf{h}} ]_{k} = [\mathbf{M} \t \mathbf{h}]_{k} \partial_{s} v_{s} - [\mathbf{M} \t \mathbf{h}]_{s}
\partial_{k} v_{s}
$$
are taken into account. Introducing for brevity notation
    $$
\mathbf{Y} = \mathbf{m} \t \mathbf{H} \equiv \mathbf{M} \t \mathbf{h} ,
    $$
we can represent the r.h.s. of Eq.~(\ref{31_07_02_10EA}) as
    $$
\frac{\partial x_{k}}{\partial a_{i}} \cdot D Y_{k}  +  Y_{s} \frac{\partial
x_{k}}{\partial a_{i}}
\partial_{k} v_{s} = \frac{\partial x_{k}}{\partial a_{i}} \cdot D Y_{k}  + Y_{s} \frac{\partial v_{s}}{\partial
a_{i}} = \frac{\partial x_{k}}{\partial a_{i}} \cdot D Y_{k}  + Y_{s}
\frac{\partial}{\partial a_{i}} (D x_{s} ) = D \left(Y_{k} \frac{\partial x_{k}}{\partial
a_{i}} \right) \, .
    $$
This proves Eq.~(\ref{31_07_02_11}).

Let us check up the integral relation (\ref{12_08_02_D_Er}). It is sufficient to prove
the differential form, namely
\begin{equation}\label{12_08_02_D_Er_PR}
D (\mathbf{M} \cdot  d \boldit{\Sigma}) = \mathbf{j} \cdot  d \boldit{\Sigma} ,
\end{equation}
where $d \boldit{\Sigma}$ is some infinitesimal oriented area moving with the fluid. It
can be presented as
\begin{equation}\label{12_08_02_D1_ER}
 d \boldit{\Sigma}  = d \mathbf{l}_{1} \t d \mathbf{l}_{2} ,
\end{equation}
where $d \mathbf{l}_{1}$, $d \mathbf{l}_{2}$ are frozen--in linear elements. Thus, $d
\mathbf{l}_{a}$, $a = 1,2$, are invariants of the $\mathbf{J}$- type and satisfy
equations
    $$
D ( d \mathbf{l}_{a}) = (d \mathbf{l}_{a} \cdot  \nabla) \mathbf{v} .
    $$
Consequently, from the recursion relation Eq.~(\ref{28_09_02_1}) it follows that $\rho d
\boldit{\Sigma}$ is $\mathbf{L}$- type invariant and hence it is governed by the dynamic
equation
    $$
D (\rho  d \boldit{\Sigma}) = - \nabla (\rho \mathbf{v} \cdot  d \boldit{\Sigma}) + \mathbf{v} \t \cu (\rho  d
\boldit{\Sigma})  ,
    $$
or in the coordinates,
\begin{equation}\label{28_09_02_2}
 D (\rho  d {\Sigma}_{i}) = - (\rho  d {\Sigma}_{k}) \partial_{i} v_{k}  \, .
\end{equation}
Now it is easy to prove relation (\ref{12_08_02_D_Er_PR}) without any restrictions for
the type of flow. Namely,
\begin{equation}\label{28_09_02_3}
\begin{split}
D (\mathbf{M} \cdot  d \boldit{\Sigma} ) = D (\mathbf{m} \cdot  \rho d \boldit{\Sigma} ) = D \mathbf{m} \cdot
\rho d \boldit{\Sigma}  +   m_{i} D (\rho d \Sigma_{i} ) = \\ = \rho d \boldit{\Sigma} \cdot  (\mathbf{m} \cdot
\nabla ) \mathbf{v}  + \mathbf{j} \cdot  d \boldit{\Sigma} - m_{i} \rho d \Sigma_{k}
\partial_{i} v_{k} =  \mathbf{j} \cdot  d \boldit{\Sigma} \, .
\end{split}
\end{equation}

In order to prove the helicity conservation, Eq.~(\ref{13_09_02_1}), let us consider some
scalar quantity of the form
    $$
Y = \mathbf{v}_{h}  \cdot \mathbf{J} ,
    $$
where $\mathbf{J}$ is some frozen--in field. Then, taking into account that
Eq.~(\ref{12_09_02_5})  for the barotropic flows can be rewritten as
    $$
D \mathbf{v}_{h} = - \nabla ( \chi - v^{2}/2)  - v_{hm} \cdot \nabla v_{m} \, , \quad
\chi \equiv \int d p /\rho ,
    $$
we obtain
    $$
D Y = D \mathbf{v}_{h} \cdot  \mathbf{J} + \mathbf{v}_{h} \cdot  D \mathbf{J} = - \nabla ( \chi - v^{2}/2) \cdot
\mathbf{J}  .
    $$
For $\mathbf{J} = \boldit{\omega}_{h}/\rho $ we proceed
    $$
D (\mathbf{v}_{h} \cdot  \boldit{\omega}_{h}/\rho) =  - \rho^{-1} \left( \nabla ( \chi - v^{2}/2) \cdot
\boldit{\omega}_{h} \right) = - \rho^{-1} \di\left(   (\chi - v^{2}/2) \boldit{\omega}_{h}  \right) \, .
    $$
Then
    $$
D (\mathbf{v}_{h} \cdot  \boldit{\omega}_{h}) = \rho D (\mathbf{v}_{h} \cdot \boldit{\omega}_{h}/\rho) +
(\mathbf{v}_{h} \cdot  \boldit{\omega}_{h}/\rho) D \rho = - \di\left(   (\chi - v^{2}/2) \boldit{\omega}_{h}
\right)  - \left(\mathbf{v}_{h} \cdot \boldit{\omega}_{h} \right) \di\mathbf{v} ,
    $$
or
\begin{equation}\label{30_09_02_A1}
\partial_{t} (\mathbf{v}_{h} \cdot  \boldit{\omega}_{h}) = - \di\mathbf{q}_{h} \, , \quad
\mathbf{q}_{h} = (\chi - v^{2}/2) \boldit{\omega}_{h} + \mathbf{v} \left(\mathbf{v}_{h}
\cdot \boldit{\omega}_{h} \right)
\end{equation}
that evidently coincides with Eq.~(\ref{13_09_02_1}).

It is noteworthy that the proof is valid for arbitrary $\mathbf{J}$-- type invariant if
the field $\rho\mathbf{J}$  is divergence--free and  the flow is barotropic:
\begin{equation}\label{30_09_02_A2}
\partial_{t} (\rho \mathbf{J} \cdot  \mathbf{v}_{h} ) = - \di\mathbf{q} \, , \quad
\mathbf{q} = (\chi - v^{2}/2) \rho \mathbf{J} + \mathbf{v} \left(\rho \mathbf{J} \cdot \mathbf{v}_{h} \right)
\quad {\mbox{for}} \quad \di(\rho\mathbf{J}) = 0 .
\end{equation}
For instance,  choosing  $\mathbf{J} = \mathbf{h}$ immediately leads to the cross--helicity invariant if one
takes into account that $\mathbf{H} \cdot \mathbf{v}_{h} = \mathbf{H} \cdot \mathbf{v}$.

\subsection*{Acknowledgment}

\frenchspacing

This work was supported by the INTAS (Grant No. 00-00292).

\medskip


\begin{thebibliography}{99}

\bibitem{Zh_Kuz_97}
V.~E.~Zakharov,  E.~A.~Kuznetsov,  {\it   Hamiltonian   formalism for nonlinear  waves}, Uspechi  Fizicheskich
Nauk (Physics Uspechi) {\bf 40}, 1087--1116 (1997).

\bibitem{GP93}
V.~P.~Goncharov,  V.~I.~Pavlov,  {\it The problems of  hydrodynamics in Hamiltonian description}, MGU
Publishers, Moskow (1993). (in Russian)

\bibitem{ZLF_92}
V.~E.~Zakharov,  V.~S.~L'vov,  G.~Falkovich, {\it Kolmogorov Spectra of Turbulence. Wave Turbulence},
Springer--Verlag, N.Y. (1992).

\bibitem{KUZ_01}
E.~A.~Kuznetsov,  {\it   Weak magnetohydrodynamic turbulence of
magnetized plasma}, Zh. Eksp. Teor. Fiz. (Sov. Phys. JETP) {\bf
120}, No~11, 1213 (2001). (in Russian)

\bibitem{Arn65}
V.~I.~Arnold, {\it  Variational  principle  for  a
three-dimensional stationary flows of ideal fluid},  {\it Prikl.
Mech.  i Mathem.} {\bf 29}, No~5, 846--851  (1965) (in Russian);
{\it Mathematical Methods of Classical Mechanics}, Springer, N.-Y.
(1978).

\bibitem{Abarb_83}
H.~D.~I.~Abarbanel, R.~Brown, Y.~M.~Yang, {\it Hamiltonian formulation of inviscid flows with free boundaries},
Phys. Fluids {\bf 31}, 2802--2809 (1988).

\bibitem{Lewis_86}
D.~Lewis, J.~Marsden, R.~Montgomery, {\it The Hamiltonian structure for hydrodynamic free boundary problems},
Physica {\bf 18D}, 391-404 (1986).

\bibitem{Vl95}
V.A. Vladimirov, H.K.~Moffatt, {\it On   General   Transformations and  variational Principles  in
Magnetohydrodynamics. Part I. Fundamental Principles.},  J.  Fl. Mech., {\bf 283}, 125--139 (1995).

\bibitem{Vl96}
V.~A.~Vladimirov, H.~K.~Moffatt, K.~I.~Ilin, {\it On   General Transformations   and variational  Principles  in
Magnetohydrodynamics.  Part II. Stability Criteria for two--dimensional Flows.}, J. Fl. Mech., {\bf 329},
187--205 (1996); {\it  Part III. Stability Criteria for Axisymmetric Flows.}, J. Plasma Phys., {\bf 57}, part 1,
89--120 (1997); {\it Part 4. Generalized Isovorticity Principle for three--dimensional Flows.}, J. Fl. Mech.,
{\bf 390} 127--150 (1999).

\bibitem{MCTYan_82}
S.~S.~Moiseev, R.~Z.~Sagdeev, A.~V.~Tur, V.~V.~Yanovsky, {\it
Frozen-in integrals of motion and Lagrange invariants in the
hydrodynamic models}, Zh. Eksp. Teor. Fiz. (Sov. Phys. JETP) {\bf
83}, No~1, 215--226 (1982) (in Russian); {\it Problems of the
theory of strong turbulence and topological solitons}. In the
book: "Nonlinear Phenomena in Plasma and Hydrodynamics", M., Mir
publishers, 137--182 (1986).

\bibitem{STY_90}
R.~Z.~Sagdeev,  A.~V.~Tur,  V.~V.~Yanovsky,  {\it Construction of Frozen--in Integrals, Lagrangian and
Topological Invariants in Hydrodynamical Models}, In Topological Fluid Mechanics, Cambridge Univ. Press, 421
(1990).

\bibitem{TYan_93}
A.~V.~Tur,  V.~V.~Yanovsky,  {\it Invariants in Dissipationless
Hydrodynamics media}, J. Fluid. Mech. {\bf 248}, No~1,  67--106
(1993).

\bibitem{VTY_95}
D.~V.~Volkov, A.~V.~Tur,  V.~V.~Yanovsky,  {\it Hidden supersymmetry of classical systems (hydrodynamics and
conservation laws)}, Phys.~Lett. A, {\bf 203}, 357--361  (1995).


\bibitem{KATS_02}
A.~V.~Kats, {\it Variational principles and canonical variables
for MHD flows with breaks I}, Radiofizika\&Radiostronomia, {\bf
7}, No~3, 232--245 (2002). (in Russian)

\bibitem{Woltjer58}
L.~Woltjer, {\it A theorem on force-free magnetic fields}, Proc. Nat. Acad. Sci.,  {\bf 44}, 489--491 (1958).

\bibitem{SteenbeckKrause66}
M.~Steenbeck, F.~Krause, {\it The generation of stellar and planetary magnetic fields by tubulent dynamo
action}, Z. Naturforsch.,  {\bf 21} A, 1285--1296 (1966).

\bibitem{Mof_69}
H.~K.~Moffatt, {\it The degree of knottedness of tangled vortex' lines}, J.~Fluid. Mech. {\bf 35}, 117--129
(1969).

\bibitem{Hameiri98}
E. Hameiri,  Phys. Plasmas {\bf 5}, 3270 (1998).

\bibitem{AlmaguerHameiriHerreraHolm88}
J. A. Almaguer, E. Hameiri, J. Herrera, and D. D. Holm,   Phys. Fluids {\bf 31}, 1930 (1988).

\bibitem{IlgisonisPastukhov96}
V. I. Ilgisonis and V. P. Pastukhov,  Physika Plasmy {\bf 22}, 228 (1996). (in Russian)

\bibitem{KATS_02A}
A.~V.~Kats,  {\it Variational principle in canonical variables,
Weber transformation and complete set of the local integrals of
motion for dissipation-free magnetohydrodynamics}, JETP Lett.,
{\bf 77}, No~12,  657--661 (2003); {\it Canonical description of
magnetohydrodynamic flows and integrals of motion}, arXiv:
physics/0212023 (2002)

       \bibitem{ZAK_KUZ_70}
V.~E.~Zakharov, E.~A.~Kuznetsov, {\it Variational principle and canonical variables in magnetohydrodynamics},
DAN SSSR, {\bf 194}, 1288 (1970). (in Russian)

\bibitem{Weber1868}
H.~Weber. {\it Ueber eine Transformation der hydrodynamischen Gleichungen}, J. Reine Angew. Math. {\bf 68},
286--292 (1868).


\bibitem{lamb}
H.~Lamb, {\it Hydrodynamics}, Cambridge Univ. Press (1932).

\bibitem{KK_97}
A.~V.~Kats,  V.~M.~Kontorovich,  {\it  Hamiltonian  description
of the motion of discontinuity surfaces},  Low Temp.  Phys.,  {\bf
23},  No~1, 89--95 (1997).


\bibitem{KATS_01}
A.~V.~Kats, {\it Variational principle and canonical variables in hydrodynamics with discontinuities}, Physica
D, {\bf 152-153}, 459--474 (2001).


\bibitem{Dirac64}
P.~A.~M.~Dirac, {\it Lectures on quantum mechanics}, Yeshiva Univ., N.Y. (1964).

\bibitem{Gitman86} D.~M.~Gitman, I.~V.~Tyutin, {\it Canonical quantization of the fields with constraints}, M., Nauka,
(1986). (in Russian)


\bibitem{Serrin59}
J.~Serrin. {\it Mathematical principles of classical fluid mechanics. Handbuch der Physik.} Stromungsmechanik I,
125--262. Springer (1959).


\bibitem{Salmon82}
R.~Salmon. {\it Hamiltonian  principle and Ertel's theorem}. In   Conf. Proc. Am. Inst. Phys. {\bf 88}, 127--135
(1982).

\bibitem{Gordin87}
V.~A.~Gordin, V.~I.~Petviashwili, Fiz. Plasmi {\bf 13}, 509 (1987). (in Russian)




\end{thebibliography}
\end{document}